\documentclass[conference]{IEEEtran}
\IEEEoverridecommandlockouts
\usepackage{cite}
\usepackage{amsmath,amssymb,amsfonts}
\usepackage{listings} 
\usepackage{algorithmic}
\usepackage{graphicx}
\usepackage{textcomp}
\usepackage{xcolor}
\usepackage{tabularray}
\usepackage[colorlinks=true,linkcolor=black,anchorcolor=black,citecolor=black,filecolor=black,menucolor=black,runcolor=black,urlcolor=black]{hyperref}
\usepackage{color}
\usepackage{array}
\usepackage{multirow}
\usepackage{ragged2e}
\usepackage{scalerel}
\usepackage{tikz}
\usetikzlibrary{svg.path}

\definecolor{orcidlogocol}{HTML}{A6CE39}
\tikzset{
  orcidlogo/.pic={
    \fill[orcidlogocol] svg{M256,128c0,70.7-57.3,128-128,128C57.3,256,0,198.7,0,128C0,57.3,57.3,0,128,0C198.7,0,256,57.3,256,128z};
    \fill[white] svg{M86.3,186.2H70.9V79.1h15.4v48.4V186.2z}
                 svg{M108.9,79.1h41.6c39.6,0,57,28.3,57,53.6c0,27.5-21.5,53.6-56.8,53.6h-41.8V79.1z M124.3,172.4h24.5c34.9,0,42.9-26.5,42.9-39.7c0-21.5-13.7-39.7-43.7-39.7h-23.7V172.4z}
                 svg{M88.7,56.8c0,5.5-4.5,10.1-10.1,10.1c-5.6,0-10.1-4.6-10.1-10.1c0-5.6,4.5-10.1,10.1-10.1C84.2,46.7,88.7,51.3,88.7,56.8z};
  }
}

\newcommand\orcidicon[1]{\href{https://orcid.org/#1}{\mbox{\scalerel*{
\begin{tikzpicture}[yscale=-1,transform shape]
\pic{orcidlogo};
\end{tikzpicture}
}{|}}}}

\usepackage{hyperref} 

\def\BibTeX{{\rm B\kern-.05em{\sc i\kern-.025em b}\kern-.08em
    T\kern-.1667em\lower.7ex\hbox{E}\kern-.125emX}}

    \makeatletter
    \newcommand{\linebreakand}{%
      \end{@IEEEauthorhalign}
      \hfill\mbox{}\par
      \mbox{}\hfill\begin{@IEEEauthorhalign}
    }
    \makeatother


\begin{document}

\title{Dynamic Power Control in a Hardware Neural Network with Error-Configurable MAC Units}

\author{
\centering
\IEEEauthorblockN{M. Ghaderi~\textsuperscript{\orcidicon{0009-0008-7675-055X}}\,A. Delavari~\textsuperscript{\orcidicon{0009-0006-6350-0055}}\, F. Ghoreishy~\textsuperscript{\orcidicon{0009-0007-7007-3386}}\, S. Mirzakuchaki~\textsuperscript{\orcidicon{0000-0003-0232-9267}}}
\IEEEauthorblockA{\textit{School of Electrical Engineering} \\
\textit{Iran University of Science and Technology}\\
Tehran, Iran \\
\{maede\_ghaderi, arvin\_delavari, faraz\_ghoreishy\}@elec.iust.ac.ir, \{m\_kuchaki\}@iust.ac.ir}
}

\maketitle

\begin{abstract}
Multi-Layer Perceptrons (MLP) are powerful tools for representing complex, non-linear relationships, making them essential for diverse machine learning and AI applications. Efficient hardware implementation of MLPs can be achieved through many hardware and architectural design techniques. These networks excel at predictive modeling and classification tasks like image classification, making them a popular choice. Approximate computing techniques are increasingly used to optimize critical path delay, area, power, and overall hardware efficiency in high-performance computing systems through controlled error and related trade-offs. This study proposes a hardware MLP neural network implemented in 45nm CMOS technology, in which MAC units of the neurons incorporate error and power controllable approximate multipliers for classification of the MNIST dataset. The optimized network consists of 10 neurons within the hidden layers, occupying 0.026mm² of area, with 5.55mW at 100MHz frequency in accurate mode and 4.81mW in lowest accuracy mode. The experiments indicate that the proposed design achieves a maximum rate of 13.33\% decrease overall and 24.78\% in each neuron’s power consumption with only a 0.92\% decrease in accuracy in comparison with accurate circuit.
\end{abstract}


\section{Introduction}

Multi-Layer Perceptron (MLP) neural networks are highly effective solutions for image classification tasks. Image classification techniques simplify visual recognition by enabling the identification of objects, faces, and even medical conditions from imaging data. The use of MLP networks for image classification enhances efficiency, accuracy, and the ability to process large-scale datasets \cite{ref-1} that would be beyond human capabilities. These Perceptrons are designed in order to learn to recognize complex patterns in data, as they are made up of interconnected nodes, similar to the neurons in a human brain. By adjusting the strengths of these connections during training, MLPs can excel at tasks like image classification, speech recognition, and financial forecasting \cite{ref-2}. The multilayered structure of MLPs allows them to capture intricate relationships within data, making them a powerful machine learning tool across various applications.

Recently there have been many investigations on hardware implementations of neural networks and AI accelerators. The primary motivation for hardware implementation of neural networks is to achieve improved performance compared to general-purpose CPUs. Neural networks, particularly the computationally intensive parts like matrix multiplications and convolutions, can be highly parallelized. FPGAs and ASICs can be designed with a large number of processing elements that can perform these operations concurrently, leading to performance improvements over CPU execution. Specialized hardware designs can be tailored to the requirements of neural network computations, reducing redundant operations and memory accesses. This can result in lower power consumption compared to general-purpose processors which is important for mobile, embedded, and edge computing applications. Integration of processing elements in these designs can minimize data transfer time, leading to lower end-to-end latency for neural network inference, which is crucial for real-time applications. Hardware implementations can scale more easily to handle larger network models and datasets compared to software-based approaches which may be limited by the memory and computational capacity of processors. The high performance and specialized nature of neural network hardware implementations make them well-suited for deployment in edge devices. Recent trends indicate there have been many progress in terms of hardware neural network design. Authors in \cite{ref-3} presented a novel hardware architecture for Feed-Forward NNs that employs two multiplexed physical layers to facilitate parallel computation. The design focuses on optimizing the architecture based on the largest layer’s neuron count, allowing it to adapt to different network sizes while minimizing execution clock cycles. A novel composite hardware architecture for CNN accelerators using FPGA that addresses throughput limitations through an optimized mapping mechanism and efficient data supply strategies was also presented in \cite{ref-4}. In \cite{ref-5}, a method for implementing DNNs on FPGA chips that leverages stochastic computing while addressing hardware constraints is presented, showcasing the effectiveness in an image processing application. Additionally, \cite{ref-6} introduces an AIoT-empowered edge-cloud collaborative computing system designed to enhance the efficiency in face detection and recognition through the use of energy-efficient FPGA-based CNN accelerators.

Nevertheless, even with specialized hardware design for the mentioned tasks, an overhead of area and power consumption may still be forced into the system due to the huge computational workload or large number of processing elements in neural networks. To address this issue, approximate computing techniques can be employed to enhance the efficiency of MLP neural networks. Approximations are allowed in applications with inherent resilient to faults injected in the systems such as image processing \cite{ref-7}, search engines \cite{ref-8} and neural networks which is the case in this research. By incorporating approximate multipliers with configurable error and power in the multiply-accumulate (MAC) units of the neurons, it is possible to reduce the hardware area and control power consumption considerably, while retaining accuracy at a reasonable level. Main contributions of this study are as follows:

\begin{itemize}
\item Design of an optimized MLP neural network for MNIST dataset classification.
\item Design and integration of an approximate multiplier with different error levels and controllable power within the MAC units of the neurons.
\item Analysis on the resulted data of the experiment in terms of hardware efficiency, overall performance, and achieved accuracy.
\end{itemize}

The remainder of this paper is structured as follows: Section II presents an overview of related and previous literature. Section III provides detailed explanations about the architecture of the proposed design. Section IV analyzes the efficiency in terms of accuracy and hardware of the proposed design with experimental results, and finally, the paper is concluded in Section V.

\section{Related Works}

Investigations were conducted on the effectiveness of approximate computing techniques in neural networks with different architectures. Approximate computing aims to trade off a small amount of accuracy for significant gains in efficiency, such as reduced energy consumption or faster processing. Approximation techniques have also been utilized in general-purpose processing cores, particularly in embedded processors \cite{ref-9}, which often operate under stringent power constraints. The authors in \cite{ref-10} proposed an adaptive solution that considers energy constraints in deep learning applications. By selectively applying approximate computing techniques, this adaptive approach can optimize the energy-accuracy tradeoff for deep neural networks. As Artificial Neural Networks (ANNs) have a wide range of applications that are inherently error-tolerant, \cite{ref-11} proposes an approximate computing framework that achieves energy savings by approximating computation and memory accesses of less critical neurons. 

Convolutional Neural Networks (CNNs) have been targeted for applying approximation techniques to present hardware-efficient structures. Some researchers in \cite{ref-12} explored techniques to make CNN architectures more efficient by approximating certain computations without significantly degrading accuracy. CNNs have been widely used for tasks like image and speech recognition due to their ability to automatically extract relevant features from the input data. \cite{ref-13} presented a reconfigurable architecture that utilizes approximate multipliers in the tensor multiplication operations of a CNN-based speech recognition system. Approximate multipliers can provide significant energy savings by trading off a small amount of precision. In \cite{ref-14}, an approximate array multiplier utilizing carry-disregarding techniques was employed in a CNN-based image classification setup. By approximating the multiplier, this approach aimed to improve the energy efficiency of the CNN model without greatly impacting its classification accuracy.

Different neural network solutions such as Support Vector Machines (SVM), Multi-Layer Perceptrons (MLP), and Deep Convolutional Neural Networks (CNN) have been used for classification of the MNIST dataset, which consists of images of handwritten digits. \cite{ref-15} explored and evaluated the performance of these various neural network architectures on the MNIST dataset. This study will also evaluate the MLPs in an image classification application using the same dataset, with the implementation of dynamic approximation control through approximate multipliers in MAC units within the neurons. The main goal of the project is to showcase the effectiveness of approximate computing techniques by measuring the power consumption in every dynamic configuration of the proposed design and evaluating the final classification accuracy of the network at different error tolerance levels.

\section{Proposed Neural Networks’s Architecture}

\begin{figure}[htbp]
\centerline{\includegraphics[width=0.35\textwidth]{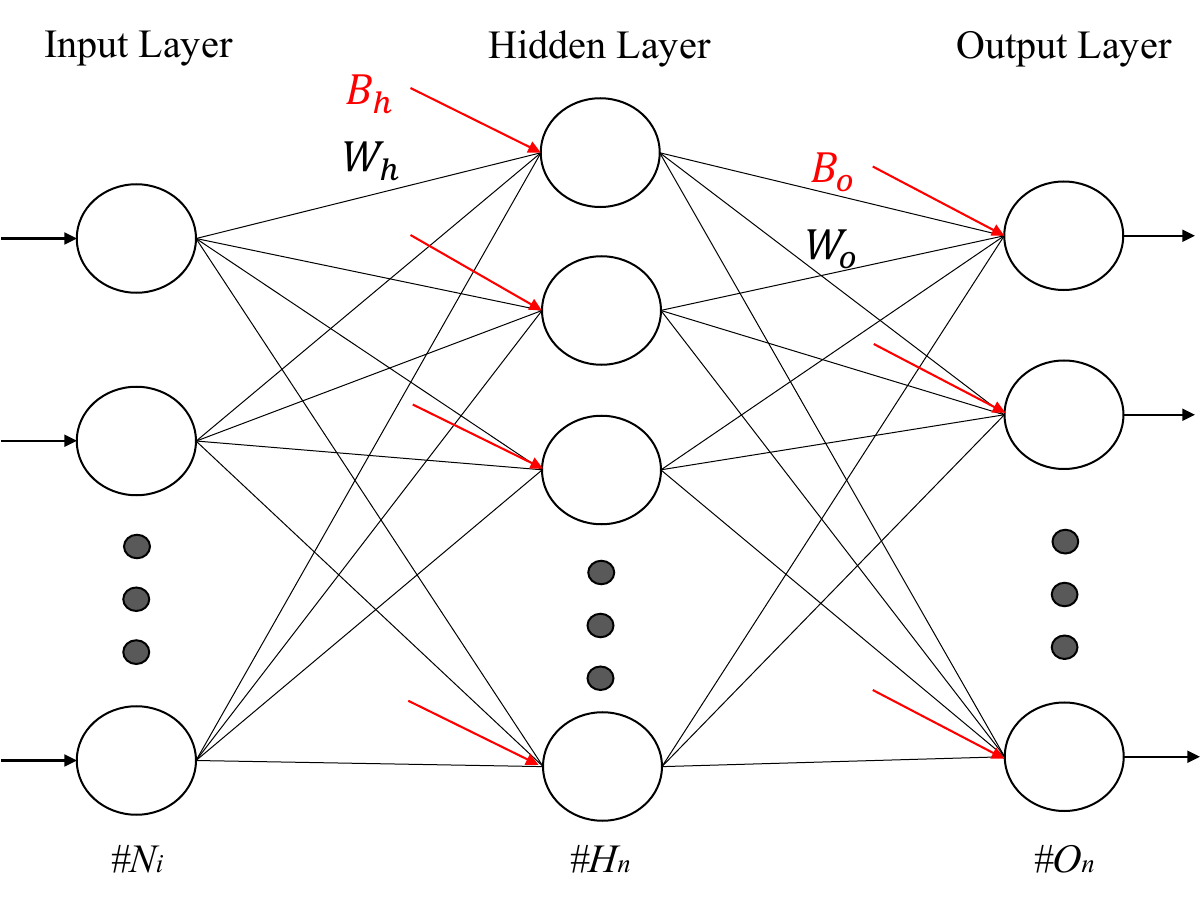}}
\caption{Proposed fully connected MLP neural network (\textit{$B_{layer}$} are for Bias factors and \textit{$W_{layer}$} are for Weights).}
\label{fig-1}
\end{figure}

In this study, a simple MLP neural network with one hidden layer was implemented, resulting in a total of three layers: an input layer, a hidden layer, and an output layer. The input layer is composed of 62 nodes, corresponding to the input features of MNIST dataset which are reduced from 748 in order to have a more hardware-efficient design. It is important to note that feature reduction algorithms can reduce the accuracy of neural networks, but for hardware implementations due to area and power constraints they are highly utilized for inputs. As illustrated in Fig.~\ref{fig-1}, the hidden layer comprises 30 neurons, where feature extraction is performed and output layer of the network consists of 10 neurons.

All connections in the proposed MLP network are fully connected. This means that each neuron in a given layer \textit{i} is connected to every neuron in the subsequent layer \textit{i – 1}, with the output of each layer being passed to the next layer with a specified weight. All weights and bias values of the trained model are kept in memory and will enter the network through controller’s given signals.

\subsection{Error Controllable MAC Unit structure}\label{A}
MAC units are crucial in neural networks as they perform the essential operations of multiplication and accumulation, which are vital for computing weighted sums. In this design, input values, weights, and biases are represented in signed magnitude format, with the Most Significant Bit (MSB) indicating the sign. If the MSB is 0, the number is positive; if it is 1, the number is negative. This representation is particularly useful for neural networks, where both positive and negative values are common.

The MAC unit employs XOR operations to manage the signs of the resulting products from the multiplication of weights and inputs. The XOR gate takes two inputs: the sign of the weight and the sign of the input. If either input is negative (MSB = 1), the output will be negative (result = 1 in terms of sign), while if both are positive (MSB = 0), the result is positive (result = 0). The rest of the operations are all done utilizing unsigned arithmetic since the sign bit is handled elsewhere.

All numbers processed in the MAC unit are 8-bit, consisting of 7 bits for the value and 1 sign bit. The result of each multiplication between an input and a weight is 14 bits wide (15 bits including the sign bit). Since the input layer consists of 62 neurons and the neural network is fully connected, 62 of these 15-bit numbers need to be summed together. This accumulation results in a 21-bit output from the MAC unit, providing sufficient range to avoid overflow when summing multiple inputs, even with high values from multiple input combinations.

\begin{figure}[htbp]
\includegraphics[width=0.5\textwidth]{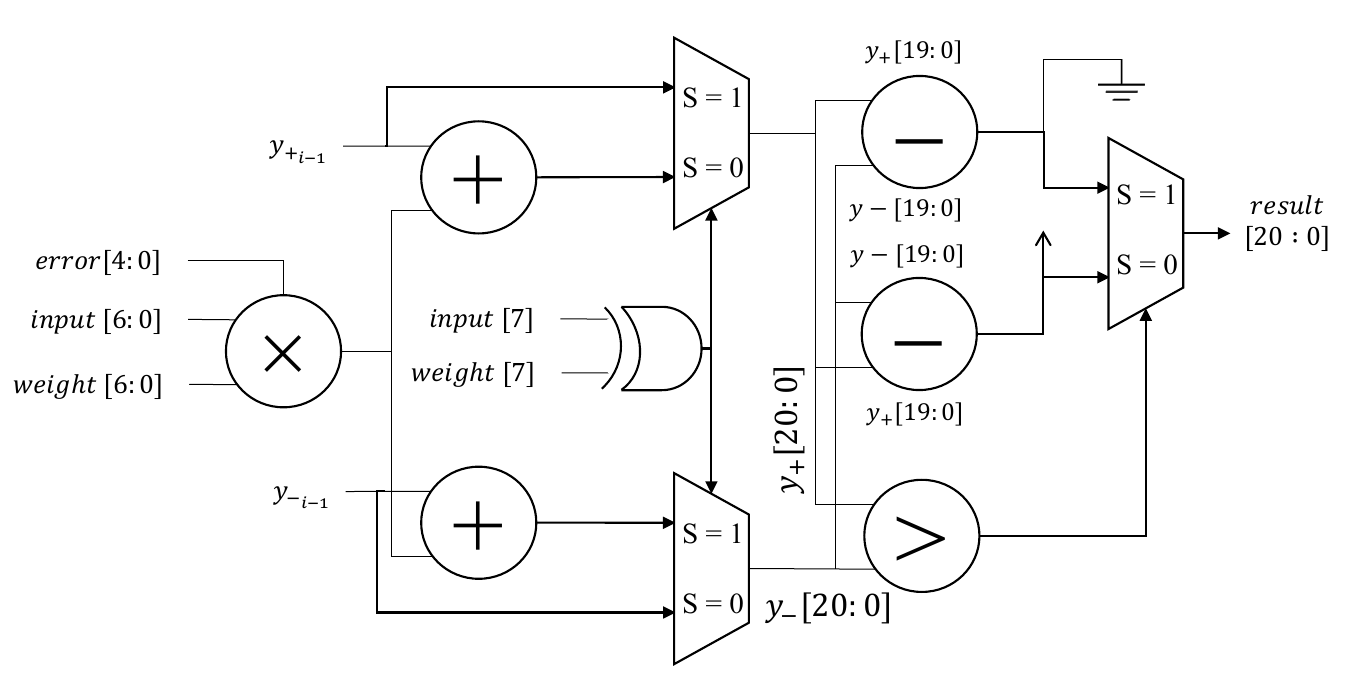}
\caption{Proposed MAC unit with error controllable approximate multiplier circuit}
\label{fig-2}
\end{figure}

Fig.~\ref{fig-2} illustrates the various components of the MAC unit, including the multiplier, which is responsible for multiplying the input and weight values, and the adder, which accumulates the results of multiple multiplications to compute the final output. Additionally, a subtractor handles cases where one of the operands is negative, while comparison logic is used to determine the final sign of the accumulated result.

The MAC unit integrated within the neurons are equipped with an approximate multiplier with dynamic error control feature. The multiplier has an error-control signal as an input in order to control output error. There are several error metrics in approximate arithmetic circuits for comparing efficiency of the designs in terms of accuracy, such as Error Rate (ER), Mean Relative Error Distance (MRED) and Normalized Mean Error Distance (NMED) \cite{ref-16} \cite{ref-17}. Table I shows average, minimum and maximum of the resulted ER, MRED and NMED of the proposed design in all 32 configurations. The proposed multiplier also has an accurate configuration in which no error is applied in the system by the arithmetic circuits. The error controllable multiplier is the key solution to dynamic power control and saving feature of the neural network. Accurate mode (configuration zero) is excluded from values in Table \ref{table-1}.

    \begin{table}[htbp]
    \caption{Accuracy Efficiency Criteria for Proposed Approximate Multiplier}
    \begin{center}
    
    \noindent\begin{minipage}{0.95\linewidth}
    \centering
    \begin{tblr}{
      width = \linewidth,
      colspec = {Q[100]Q[100]Q[100]Q[100]Q[100]Q[100]},
      cells = {c},
      cell{1}{1} = {c=2}{0.250\linewidth},
      cell{1}{3} = {c=2}{0.250\linewidth},
      cell{1}{5} = {c=2}{0.250\linewidth},
      cell{4}{1} = {c=6}{0.75\linewidth},
      cell{5}{1} = {c=2}{0.250\linewidth},
      cell{5}{3} = {c=2}{0.250\linewidth},
      cell{5}{5} = {c=2}{0.250\linewidth},
      hlines,
      vlines,
    }
    ER [\%] &  & MRED [\%] &  & NMED [\%] & \\
    MIN & MAX & MIN & MAX & MIN & MAX\\
    9.9609 & 61.8255 & 0.0548 & 3.6840 & 0.0028 & 0.3643\\
    Average (32 configurations) &  &  &  &  & \\
    43.556 &  & 2.125 &  & 0.224 & 
    \end{tblr}
    \end{minipage}

    \label{table-1}
    \end{center}
    \end{table}

\subsection{Neurons’ structure}
As shown in (\ref{equation-1}) neurons process input data to produce an output by applying weights and summing them, which is done utilizing MAC units, adding the bias terms, and passing the result through an activation function. In (\ref{equation-1}), \textit{$W_{ij}$} represents defined weights between \textit{$i^{th}$} output of previous layer and \textit{$j^{th}$} nodes of following layer, and \textit{$b_{j}$} denotes the \textit{$j^{th}$} respective biases. According to the number of inputs (\textit{$\#N_{i}$}) in this setup, and the number of neurons in the hidden layer (\textit{$\#H_{n}$}), the network will have \textit{$\#N_{i}\times\#H_{n}$} hidden layer weights. Additionally, considering \textit{$\#O_{n}$} neurons in output layer, there are \textit{$\#H_{n}\times\#O_{n}$} output layer weights. Also, the number of needful biases in hidden and output layers are equal to \textit{$\#H_{n}$} and \textit{$\#O_{n}$}, respectively.

    \begin{equation}\label{equation-1}
    y_i=(\sum_{i=0}^{n}W_{ij}\times x_i + b_j)
    \end{equation}

\begin{figure}[htbp]
\centerline{\includegraphics[width=0.5\textwidth]{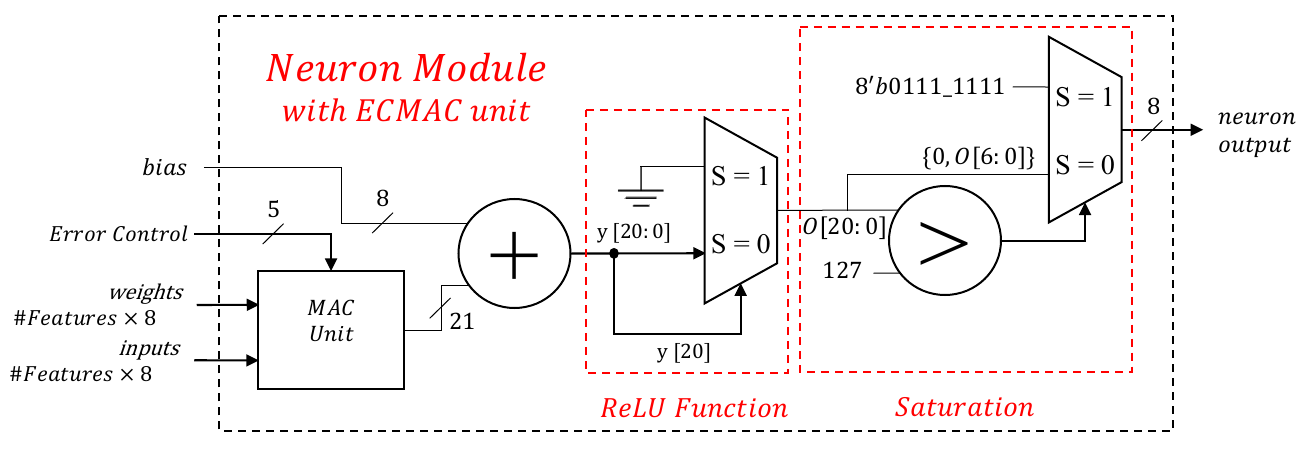}}
\caption{Structure of neuron module in the proposed MLP neural network with error controllable MAC units}
\label{fig-3}
\end{figure}

In the proposed design, each neuron contains a MAC unit, an adder to include the bias, an activation function (ReLU in this paper), and a saturation section to limit the activation output from 21 bits to 8 bits, ensuring compatibility with the input requirements of the subsequent layer. The block diagram of neurons implemented in this MLP network is shown in Fig.~\ref{fig-3}.

Due to necessity of activation functions for their ability to introduce non-linearity into neural networks, Rectified Linear Unit (ReLU) function, which is a non-linear function, is implemented in the proposed design. ReLU is described as (\ref{equation-2}) allowing the network to learn and capture intricate patterns by activating neurons only when their input is positive. ReLU is used in hardware neural networks more frequently than software approaches, due to its simplicity in implementation.

    \begin{equation}\label{equation-2}
    ReLU(x) = max (0, x)
    \end{equation}

\subsection{Datapath and overall structure}
\begin{figure}[htbp]
\centerline{\includegraphics[width=0.49\textwidth]{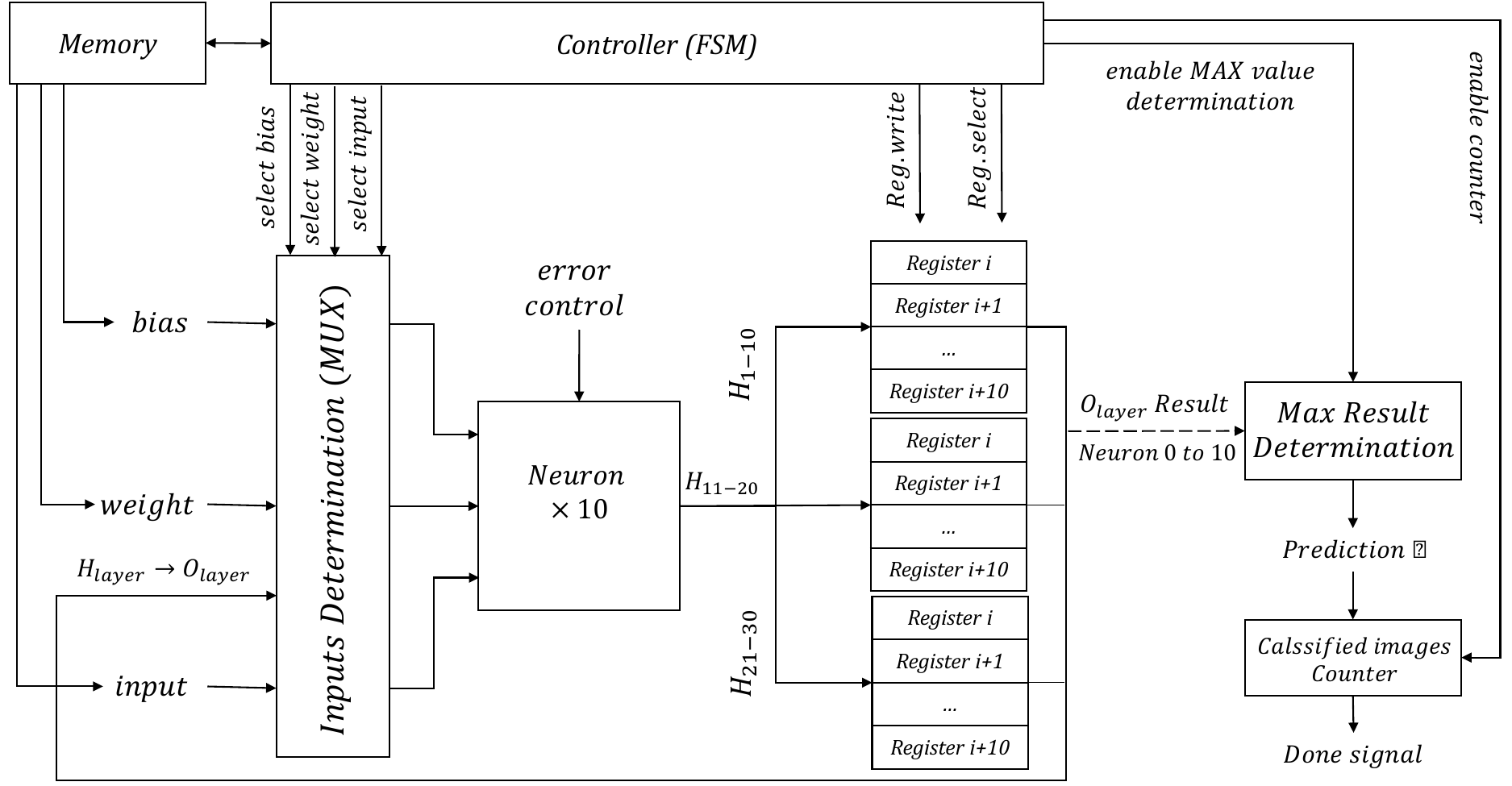}}
\caption{Datapath of the implemented MLP neural network}
\label{fig-4}
\end{figure}

Considering both hidden and output layers, \textit{$\#H_{n}+\#O_{n}$} neurons are totally needed. However, in order to save resources only 10 hardware neurons are integrated, working in 4 separate cycles to execute all neurons’ operations. To handle the multicycle datapath, a controller is designed, which will be discussed later. As is shown in Fig.~\ref{fig-4}, the input values, weights, and biases are chosen within multiplexers based on their relevant field with selection signals. For each of the 10 neurons in the hidden layer, 10 registers with 8-bit width are incorporated to store the results, which then serve as input for the output layer. Finally, by evaluating results of output layer, the maximum output value, as shown in Fig.~\ref{fig-4}, is acquired and the prediction label emerges.

This approach ensures efficient use of hardware resources while maintaining the accuracy and functionality of the neural network. The design effectively manages the data flow and computation across multiple cycles, enabling network to process inputs and generate predictions as required.

\subsection{Controller}
To manage input data, weights and biases, a finite state machine with 5 states is designed. This state machine coordinates a multi-step process by flow control of the operations, including reading inputs, selecting computation parameters (weights and biases), loading registers, and enabling counters. It transitions through a series of states, ultimately signaling when the entire process is complete.

\begin{itemize}
\item State 0: The state machine starts by enabling the reading of memory inputs and selecting the input source. It sets the weight selections and bias selection to 0 to choose the proper weights and biases which are related to first 10 neurons in hidden layer, and the results of all neuron in this state will be stored into registers. The system then transitions to State 1.
\item State 1: In this state, the machine continues reading memory inputs and selecting the input source. The weight and bias selections are set to 1, which select the parameters of the second set of 10 hidden layer neurons, and the outputs from all neurons in the same state will be saved into registers. The state machine then moves to State 2.
\item State 2: The operations of reading memory inputs and selecting the input source continues. The weight and bias selections are updated to 2 to fetch the last 10 neurons’ weights and biases in hidden layer and the results of each neuron in the same state will be stored in registers. The machine transitions to State 3.
\item Sate 3: The state machine updates the weight and bias selections to 3 in order to choose output layer’s variable. Then, the circuit for maximum value determination is enabled to obtain the predicted label for input image. Furthermore, classified image counter is activated, and if the counter value is less than number of input images in dataset, the state machine loops back to State 0; otherwise, it progresses to State 4.
\item State 4: In the following state, controller produces a signal for concluding operations when all images of the dataset are classified.
\end{itemize}

\section{Experimental Results}

The proposed neural network was implemented in 45nm CMOS technology utilizing Synopsys Design Compiler software. Power analysis was accomplished by implying related switching activity files to the proposed MLP network, which was designed using Verilog HDL. The dataset was loaded into an external memory space within the testbench, which served as input for the proposed MLP. An integer variable within the testbench was utilized to count the number of correct results, as indicated by the number of high states of the prediction completion signal during each interval. The accuracy of the proposed MLP is measured by dividing the count of correctly predicted results by the total number of images. As depicted in Fig.~\ref{fig-4}, the FSM is tasked with managing the sequence of input data (in each clock cycle) sourced from memory and identifying the type of input, whether it is a weight, bias, or standard input data. The proposed error and power controllable multiplier in the MAC units have 32 configurations. Proposed design has reached an area of 26084µm², operating in a frequency range of 100MHz to 330MHz. All power results in this study were measured in the equal setup, where design is working at 100MHz frequency at 1.1V. The ER and MRED of the proposed multiplier vary from 0.00\% to 61.82\% and 0.00\% to 3.68\% regarding the dynamic error control feature of the design. The average ER of the design is equal to 43.55\% and average MRED is 2.12\%. The average Normalized Mean Error Distance (NMED) is 0.22. The multiplier circuit was carefully designed in order to provide the maximum range of power saving in different configurations, and minimum error injection to the system. 

The proposed design was evaluated using all 32 configurations of the MAC units, testing each configuration across every set of 10 neurons. The results demonstrated success in terms of power efficiency and final accuracy. Accuracy of the MLP is extracted from the ratio of correctly classified images to the number of all images in the preferred dataset. The maximum accuracy reached through this system was 89.67\% (according to hardware optimizations and approximation in design process). The lowest accuracy reached through approximate mode of the MAC units in the most inaccurate configuration was 88.75\% which shows that accuracy has fallen less than 1\% in the worst case. The average accuracy of all 32 configurations is 89.11\% which has a difference of only 0.56\% in comparison with accurate form which indicates reasonable results. But the most important achievement in this study is the power saved in configurations adhering approximation in MAC units.

\begin{figure}[htbp]
\centerline{\includegraphics[width=0.5\textwidth]{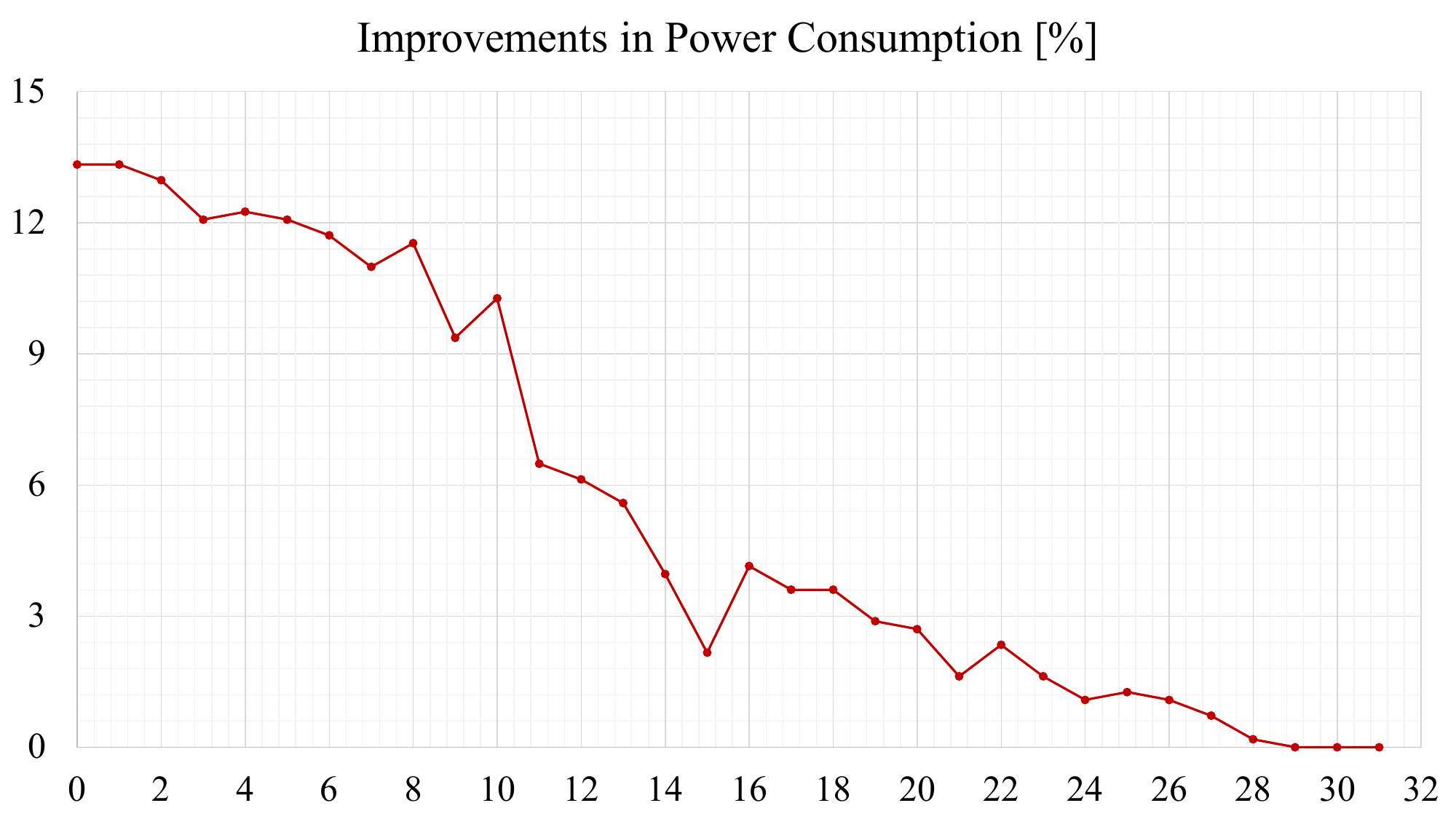}}
\caption{Improvements in overall power consumption all 32 dynamic configurations of MAC units (neurons)}
\label{fig-5}
\end{figure}

\begin{figure}[htbp]
\centerline{\includegraphics[width=0.5\textwidth]{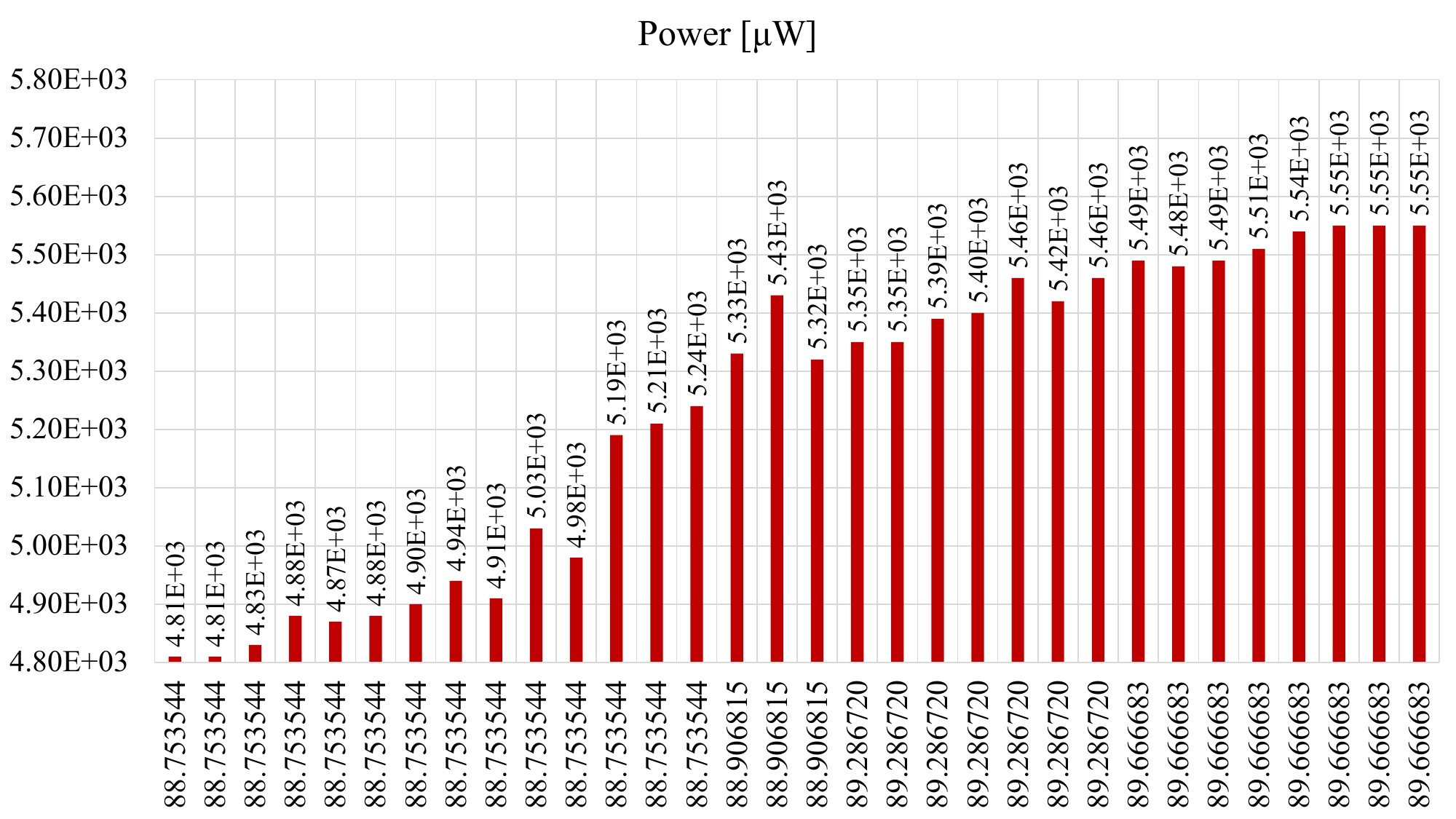}}
\caption{Power consumption based on accuracy of the MLP neural network in all 32 configurations of MAC units (neurons)}
\label{fig-6}
\end{figure}

Fig.~\ref{fig-5} presents percentage of improvement in terms of power consumption, as Fig.~\ref{fig-6} shows the power consumed in every configuration. 

Fig.~\ref{fig-6} is presenting power consumption based on accuracy in the respected error level. In the worst accuracy configuration of the MAC unit in which the multiplier has an ER of 61.82\% and MRED of 3.68\%, the accuracy is 88.75\% where the power consumed in all of the network is equal to 4.81mW.

\begin{figure}[htbp]
\centerline{\includegraphics[width=0.5\textwidth]{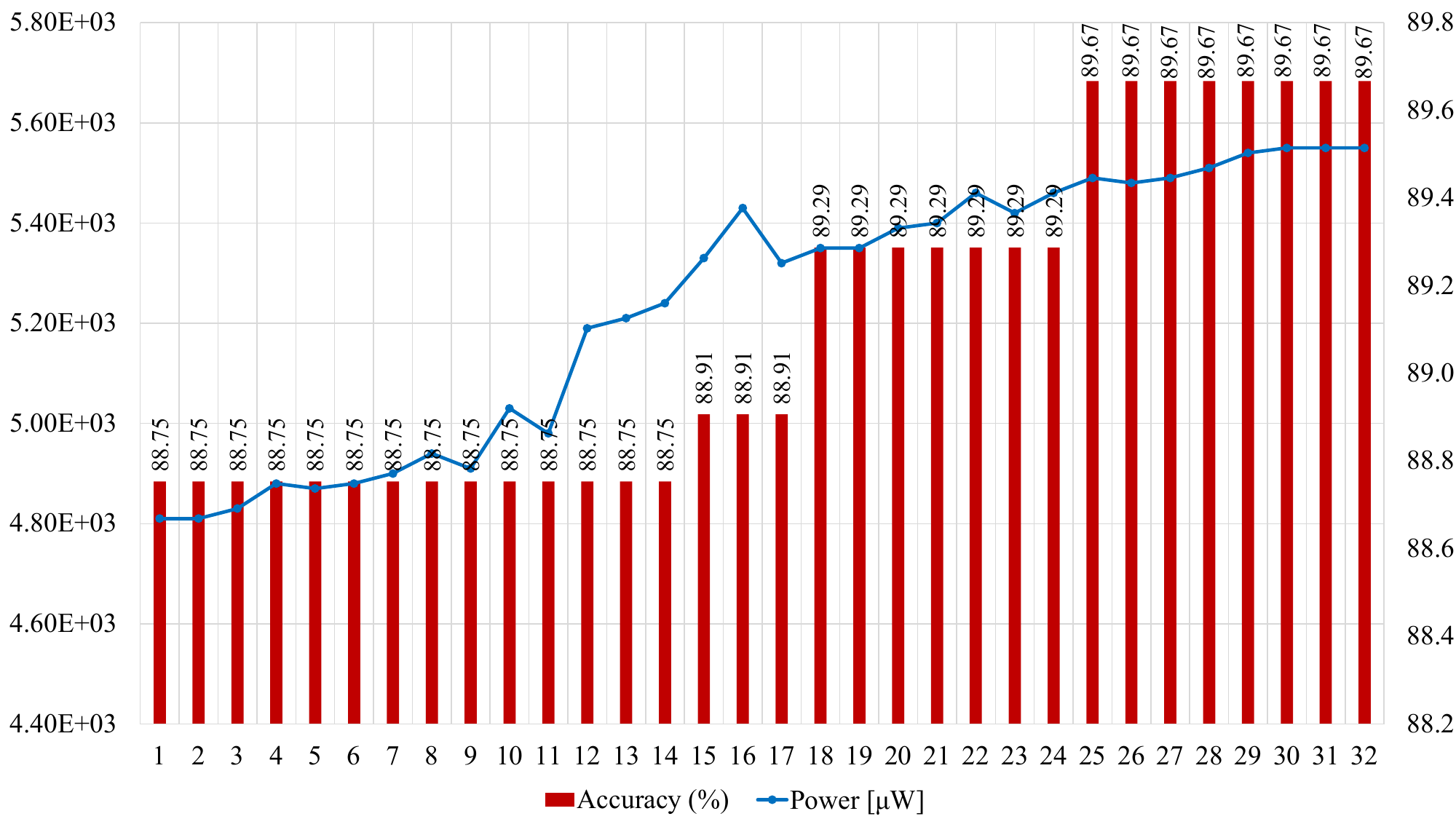}}
\caption{Neural network accuracy trade-off with overall power consumption of implemented hardware}
\label{fig-7}
\end{figure}

Accurate mode of MAC units resulted in a power consumption of 5.55mW. In fact, the maximum power saved in the system is 740µW which is equal to 13.33\% improvement in all of the neural network, 44.36\% in each MAC unit and 24.78\% in every neuron. The average improvement of all 32 configurations is equal to 5.84\% in total power consumption which is nearly 324µW. Average improvements in power saved by MAC units is 40.89\% and in neurons is 22.90\%. Fig.~\ref{fig-7} illustrates the trade-off between accuracy and power consumption in the proposed design which has led to reasonable results, providing a promising approach for power saving in hardware neural networks.

The proposed design is highly optimized through a combination of feature reduction and resource-saving techniques. This involved reducing the number of neurons in both hidden and output layers and applying operations across multiple clock cycles, leading to a hardware efficient design. The key finding of this study is that employing circuit-level approximation and hardware-based techniques in neural networks can yield substantial power savings. Each neuron out of the 10 in the proposed MLP achieved an average 22.90\% power savings compared to the accurate version solely. Furthermore, if scaled to larger neural network architectures, the impact of these hardware approximation techniques would become even more obvious.

\section{Conclusion}

This paper introduced an MLP neural network with approximate MAC units to enhance energy efficiency for image classification tasks demonstrated on the MNIST dataset. Our design achieved a 13.33\% improvement in power consumption with only a 0.92\% drop in accuracy compared to exact computation instance in the least accurate configuration which is a remarkable result. These findings illustrate the effectiveness of approximate computing techniques in improving hardware efficiency while maintaining acceptable accuracy regarding the application. The results point out the practicality of addressing the inefficiencies seen in conventional MLP implementations by using approximation methods. The reduction in hardware area, power consumption and critical path delay is demanded for applications that require low-power and compact hardware solutions, such as portable and embedded systems.

Furthermore, the slight trade-off in accuracy is acceptable for many practical applications, especially when the benefits in terms of efficiency are considerable. The suggested accurate/approximate MLP method is a strong choice for many machine learning tasks because it can keep a reasonable degree of accuracy while enhancing hardware performance. Future work will explore the application of approximate computing solutions to other neural network architectures and datasets to validate their generalizability and effectiveness. Additionally, further optimizations could be investigated to minimize the accuracy trade-off. The integration of approximate arithmetic circuits into more complex neural networks shall pave the way for more energy-efficient AI solutions across various domains.

In conclusion, this study presents a notable direction for enhancing the efficiency of neural network hardware through the use of approximate computing techniques. By achieving a proper balance between performance and resource utilization, the proposed approach offers significant advantages for the development of efficient, high-performance machine learning and artificial intelligence systems.


\begin{thebibliography}{00}

\bibitem{ref-1} C. Xu, ``Applying MLP and CNN on Handwriting Images for Image Classification Task,'' \textit{2022 5th International Conference on Advanced Electronic Materials, Computers and Software Engineering (AEMCSE)}, Wuhan, China, 2022, pp. 830-835.

\bibitem{ref-2} R. S. Chugh, V. Bhatia, K. Khanna and V. Bhatia, ``A Comparative Analysis of Classifiers for Image Classification,'' \textit{2020 10th International Conference on Cloud Computing, Data Science \& Engineering}, Noida, India, 2020, pp. 248-253.

\bibitem{ref-3} El-Sharkawy, Mohamed et al. ``Re-configurable parallel Feed-Forward Neural Network implementation using FPGA.'' Integr. 97 (2024): 102176

\bibitem{ref-4} W. Huang et al., ``FPGA-Based High-Throughput CNN Hardware Accelerator With High Computing Resource Utilization Ratio,'' \textit{in IEEE Transactions on Neural Networks and Learning Systems}, vol. 33, no. 8, pp. 4069-4083, Aug. 2022.

\bibitem{ref-5} Nobari, Maedeh and Hadi Jahanirad. ``FPGA-based implementation of deep neural network using stochastic computing.'' Appl. Soft Comput. 137 (2023): 110166.

\bibitem{ref-6} X. Liu et al., ``Collaborative Edge Computing With FPGA-Based CNN Accelerators for Energy-Efficient and Time-Aware Face Tracking System,'' \textit{in IEEE Transactions on Computational Social Systems}, vol. 9, no. 1, pp. 252-266, Feb. 2022

\bibitem{ref-7} G. Anusha and P. Deepa, ``Design of approximate adders and multipliers for error tolerant image processing,'' \textit{Microprocessors and Microsystems}, vol. 72, p. 102940, 2020.

\bibitem{ref-8} S. Mittal, ``A survey of techniques for approximate computing,'' \textit{ACM Comput. Surv.}, vol. 48, no. 4, pp. 62:1-62:33, Mar. 2016

\bibitem{ref-9} A. Delavari, F. Ghoreishy, H. S. Shahhoseini and S. Mirzakuchaki, ``A Reconfigurable Approximate Computing RISC-V Platform for Fault-Tolerant Applications,'' \textit{2024 27th Euromicro Conference on Digital System Design (DSD)}, Paris, France, 2024, pp. 81-89.

\bibitem{ref-10} N. TaheriNejad and S. Shakibhamedan, ``Energy-aware adaptive approximate computing for deep learning applications,'' \textit{in Proc. IEEE Comput.Soc. Annu. Symp. VLSI (ISVLSI)}, Jul. 2022, p. 328.

\bibitem{ref-11} Q. Zhang, T. Wang, Y. Tian, F. Yuan and Q. Xu, ``ApproxANN: An approximate computing framework for artificial neural network,'' \textit{2015 Design, Automation \& Test in Europe Conference \& Exhibition (DATE)}, Grenoble, France, 2015, pp. 701-706.

\bibitem{ref-12} M. F. Tolba, H. Saleh, M. Al-Qutayri, A. Hroub and T. Stouraitis, ``Efficient CNN Hardware Architecture Based on Linear Approximation and Computation Reuse Technique,'' \textit{2023 International Conference on Microelectronics (ICM)}, Abu Dhabi, United Arab Emirates, 2023, pp. 7-10.

\bibitem{ref-13} J. Qian, Y. Jiang, Z. Zhang, R. Zhang, Z. Wang and B. Liu, ``Reconfigurable Approximate Multiplication Architecture for CNN-Based Speech Recognition Using Wallace Tree Tensor Multiplier Unit,'' \textit{2021 IEEE/ACM International Symposium on Nanoscale Architectures (NANOARCH)}, AB, Canada, 2021, pp. 1-6.

\bibitem{ref-14} S. Shakibhamedan, N. Amirafshar, A. S. Baroughi, H. S. Shahhoseini and N. Taherinejad, ``ACE-CNN: Approximate carry disregard multipliers for energy-efficient CNN-Based image classification,'' \textit{in IEEE Transactions on Circuits and Systems I: Regular Papers}, 2024.

\bibitem{ref-15} N. Mohamed, R. Josphineleela, S. Rakhamaji Madkar, J. Vasantha Sena, B. S. Alfurhood and B. Pant, ``The Smart Handwritten Digits Recognition Using Machine Learning Algorithm,'' \textit{2023 3rd International Conference on Advance Computing and Innovative Technologies in Engineering (ICACITE)}, Greater Noida, India, 2023, pp. 340-344.

\bibitem{ref-16} A. G. M. Strollo, E. Napoli, D. D. Caro, N. Petra, and G. D. Meo, ``Comparison and extension of approximate 4–2 compressors for lowpower approximate multipliers,'' \textit{IEEE Trans. Circuits Syst. I, Reg. Papers}, vol. 67, no. 9, pp. 3021–3034, Sep. 2020.

\bibitem{ref-17} P. Yin, C. Wang, H. Waris, W. Liu, Y. Han, and F. Lombardi, ``Design and analysis of energy-efficient dynamic range approximate logarithmic multipliers for machine learning,'' \textit{IEEE Trans. Sustain. Comput.}, vol. 6, no. 4, pp. 612–625, Oct. 2021.

\end{thebibliography}
\end{document}